\DocumentMetadata{}
\documentclass[sigconf]{acmart}
\AtBeginDocument{}

\usepackage{booktabs}
\usepackage{multirow}
\usepackage{rotating}
\usepackage{comment}
\usepackage{xcolor}
\usepackage{soul}
\usepackage{enumitem}
\usepackage[textfont=normalfont,skip=3pt]{caption}
\usepackage{anyfontsize}

\setlist[enumerate]{leftmargin=*}

\makeatletter
\@namedef{ver@lineno.sty}{9999/12/31}
\@namedef{opt@lineno.sty}{}
\makeatother

\usepackage{newfloat}
\DeclareCaptionType{example}

\usepackage{tabularx}
\usepackage[group-separator={,}]{siunitx}

\usepackage{float}
\restylefloat{table}

\copyrightyear{2025}
\acmYear{2025}
\setcopyright{rightsretained}
\acmConference[SIGCSE TS 2025]{Proceedings of the 56th ACM Technical Symposium on Computer Science Education V. 1}{February 26-March 1, 2025}{Pittsburgh, PA, USA}
\acmBooktitle{Proceedings of the 56th ACM Technical Symposium on Computer Science Education V. 1 (SIGCSE TS 2025), February 26-March 1, 2025, Pittsburgh, PA, USA}
\acmDOI{10.1145/3641554.3701862}
\acmISBN{979-8-4007-0531-1/25/02}

\usepackage{fvextra}
\usepackage{upquote}

\makeatletter
\def\PYG@reset{\let\PYG@it=\relax \let\PYG@bf=\relax%
    \let\PYG@ul=\relax \let\PYG@tc=\relax%
    \let\PYG@bc=\relax \let\PYG@ff=\relax}
\def\PYG@tok#1{\csname PYG@tok@#1\endcsname}
\def\PYG@toks#1+{\ifx\relax#1\empty\else%
    \PYG@tok{#1}\expandafter\PYG@toks\fi}
\def\PYG@do#1{\PYG@bc{\PYG@tc{\PYG@ul{%
    \PYG@it{\PYG@bf{\PYG@ff{#1}}}}}}}
\def\PYG#1#2{\PYG@reset\PYG@toks#1+\relax+\PYG@do{#2}}

\@namedef{PYG@tok@w}{\def\PYG@tc##1{\textcolor[rgb]{0.73,0.73,0.73}{##1}}}
\@namedef{PYG@tok@c}{\let\PYG@it=\textit\def\PYG@tc##1{\textcolor[rgb]{0.24,0.48,0.48}{##1}}}
\@namedef{PYG@tok@cp}{\def\PYG@tc##1{\textcolor[rgb]{0.61,0.40,0.00}{##1}}}
\@namedef{PYG@tok@k}{\let\PYG@bf=\textbf\def\PYG@tc##1{\textcolor[rgb]{0.00,0.50,0.00}{##1}}}
\@namedef{PYG@tok@kp}{\def\PYG@tc##1{\textcolor[rgb]{0.00,0.50,0.00}{##1}}}
\@namedef{PYG@tok@kt}{\def\PYG@tc##1{\textcolor[rgb]{0.69,0.00,0.25}{##1}}}
\@namedef{PYG@tok@o}{\def\PYG@tc##1{\textcolor[rgb]{0.40,0.40,0.40}{##1}}}
\@namedef{PYG@tok@ow}{\let\PYG@bf=\textbf\def\PYG@tc##1{\textcolor[rgb]{0.67,0.13,1.00}{##1}}}
\@namedef{PYG@tok@nb}{\def\PYG@tc##1{\textcolor[rgb]{0.00,0.50,0.00}{##1}}}
\@namedef{PYG@tok@nf}{\def\PYG@tc##1{\textcolor[rgb]{0.00,0.00,1.00}{##1}}}
\@namedef{PYG@tok@nc}{\let\PYG@bf=\textbf\def\PYG@tc##1{\textcolor[rgb]{0.00,0.00,1.00}{##1}}}
\@namedef{PYG@tok@nn}{\let\PYG@bf=\textbf\def\PYG@tc##1{\textcolor[rgb]{0.00,0.00,1.00}{##1}}}
\@namedef{PYG@tok@ne}{\let\PYG@bf=\textbf\def\PYG@tc##1{\textcolor[rgb]{0.80,0.25,0.22}{##1}}}
\@namedef{PYG@tok@nv}{\def\PYG@tc##1{\textcolor[rgb]{0.10,0.09,0.49}{##1}}}
\@namedef{PYG@tok@no}{\def\PYG@tc##1{\textcolor[rgb]{0.53,0.00,0.00}{##1}}}
\@namedef{PYG@tok@nl}{\def\PYG@tc##1{\textcolor[rgb]{0.46,0.46,0.00}{##1}}}
\@namedef{PYG@tok@ni}{\let\PYG@bf=\textbf\def\PYG@tc##1{\textcolor[rgb]{0.44,0.44,0.44}{##1}}}
\@namedef{PYG@tok@na}{\def\PYG@tc##1{\textcolor[rgb]{0.41,0.47,0.13}{##1}}}
\@namedef{PYG@tok@nt}{\let\PYG@bf=\textbf\def\PYG@tc##1{\textcolor[rgb]{0.00,0.50,0.00}{##1}}}
\@namedef{PYG@tok@nd}{\def\PYG@tc##1{\textcolor[rgb]{0.67,0.13,1.00}{##1}}}
\@namedef{PYG@tok@s}{\def\PYG@tc##1{\textcolor[rgb]{0.73,0.13,0.13}{##1}}}
\@namedef{PYG@tok@sd}{\let\PYG@it=\textit\def\PYG@tc##1{\textcolor[rgb]{0.73,0.13,0.13}{##1}}}
\@namedef{PYG@tok@si}{\let\PYG@bf=\textbf\def\PYG@tc##1{\textcolor[rgb]{0.64,0.35,0.47}{##1}}}
\@namedef{PYG@tok@se}{\let\PYG@bf=\textbf\def\PYG@tc##1{\textcolor[rgb]{0.67,0.36,0.12}{##1}}}
\@namedef{PYG@tok@sr}{\def\PYG@tc##1{\textcolor[rgb]{0.64,0.35,0.47}{##1}}}
\@namedef{PYG@tok@ss}{\def\PYG@tc##1{\textcolor[rgb]{0.10,0.09,0.49}{##1}}}
\@namedef{PYG@tok@sx}{\def\PYG@tc##1{\textcolor[rgb]{0.00,0.50,0.00}{##1}}}
\@namedef{PYG@tok@m}{\def\PYG@tc##1{\textcolor[rgb]{0.40,0.40,0.40}{##1}}}
\@namedef{PYG@tok@gh}{\let\PYG@bf=\textbf\def\PYG@tc##1{\textcolor[rgb]{0.00,0.00,0.50}{##1}}}
\@namedef{PYG@tok@gu}{\let\PYG@bf=\textbf\def\PYG@tc##1{\textcolor[rgb]{0.50,0.00,0.50}{##1}}}
\@namedef{PYG@tok@gd}{\def\PYG@tc##1{\textcolor[rgb]{0.63,0.00,0.00}{##1}}}
\@namedef{PYG@tok@gi}{\def\PYG@tc##1{\textcolor[rgb]{0.00,0.52,0.00}{##1}}}
\@namedef{PYG@tok@gr}{\def\PYG@tc##1{\textcolor[rgb]{0.89,0.00,0.00}{##1}}}
\@namedef{PYG@tok@ge}{\let\PYG@it=\textit}
\@namedef{PYG@tok@gs}{\let\PYG@bf=\textbf}
\@namedef{PYG@tok@ges}{\let\PYG@bf=\textbf\let\PYG@it=\textit}
\@namedef{PYG@tok@gp}{\let\PYG@bf=\textbf\def\PYG@tc##1{\textcolor[rgb]{0.00,0.00,0.50}{##1}}}
\@namedef{PYG@tok@go}{\def\PYG@tc##1{\textcolor[rgb]{0.44,0.44,0.44}{##1}}}
\@namedef{PYG@tok@gt}{\def\PYG@tc##1{\textcolor[rgb]{0.00,0.27,0.87}{##1}}}
\@namedef{PYG@tok@err}{\def\PYG@bc##1{{\setlength{\fboxsep}{\string -\fboxrule}\fcolorbox[rgb]{1.00,0.00,0.00}{1,1,1}{\strut ##1}}}}
\@namedef{PYG@tok@kc}{\let\PYG@bf=\textbf\def\PYG@tc##1{\textcolor[rgb]{0.00,0.50,0.00}{##1}}}
\@namedef{PYG@tok@kd}{\let\PYG@bf=\textbf\def\PYG@tc##1{\textcolor[rgb]{0.00,0.50,0.00}{##1}}}
\@namedef{PYG@tok@kn}{\let\PYG@bf=\textbf\def\PYG@tc##1{\textcolor[rgb]{0.00,0.50,0.00}{##1}}}
\@namedef{PYG@tok@kr}{\let\PYG@bf=\textbf\def\PYG@tc##1{\textcolor[rgb]{0.00,0.50,0.00}{##1}}}
\@namedef{PYG@tok@bp}{\def\PYG@tc##1{\textcolor[rgb]{0.00,0.50,0.00}{##1}}}
\@namedef{PYG@tok@fm}{\def\PYG@tc##1{\textcolor[rgb]{0.00,0.00,1.00}{##1}}}
\@namedef{PYG@tok@vc}{\def\PYG@tc##1{\textcolor[rgb]{0.10,0.09,0.49}{##1}}}
\@namedef{PYG@tok@vg}{\def\PYG@tc##1{\textcolor[rgb]{0.10,0.09,0.49}{##1}}}
\@namedef{PYG@tok@vi}{\def\PYG@tc##1{\textcolor[rgb]{0.10,0.09,0.49}{##1}}}
\@namedef{PYG@tok@vm}{\def\PYG@tc##1{\textcolor[rgb]{0.10,0.09,0.49}{##1}}}
\@namedef{PYG@tok@sa}{\def\PYG@tc##1{\textcolor[rgb]{0.73,0.13,0.13}{##1}}}
\@namedef{PYG@tok@sb}{\def\PYG@tc##1{\textcolor[rgb]{0.73,0.13,0.13}{##1}}}
\@namedef{PYG@tok@sc}{\def\PYG@tc##1{\textcolor[rgb]{0.73,0.13,0.13}{##1}}}
\@namedef{PYG@tok@dl}{\def\PYG@tc##1{\textcolor[rgb]{0.73,0.13,0.13}{##1}}}
\@namedef{PYG@tok@s2}{\def\PYG@tc##1{\textcolor[rgb]{0.73,0.13,0.13}{##1}}}
\@namedef{PYG@tok@sh}{\def\PYG@tc##1{\textcolor[rgb]{0.73,0.13,0.13}{##1}}}
\@namedef{PYG@tok@s1}{\def\PYG@tc##1{\textcolor[rgb]{0.73,0.13,0.13}{##1}}}
\@namedef{PYG@tok@mb}{\def\PYG@tc##1{\textcolor[rgb]{0.40,0.40,0.40}{##1}}}
\@namedef{PYG@tok@mf}{\def\PYG@tc##1{\textcolor[rgb]{0.40,0.40,0.40}{##1}}}
\@namedef{PYG@tok@mh}{\def\PYG@tc##1{\textcolor[rgb]{0.40,0.40,0.40}{##1}}}
\@namedef{PYG@tok@mi}{\def\PYG@tc##1{\textcolor[rgb]{0.40,0.40,0.40}{##1}}}
\@namedef{PYG@tok@il}{\def\PYG@tc##1{\textcolor[rgb]{0.40,0.40,0.40}{##1}}}
\@namedef{PYG@tok@mo}{\def\PYG@tc##1{\textcolor[rgb]{0.40,0.40,0.40}{##1}}}
\@namedef{PYG@tok@ch}{\let\PYG@it=\textit\def\PYG@tc##1{\textcolor[rgb]{0.24,0.48,0.48}{##1}}}
\@namedef{PYG@tok@cm}{\let\PYG@it=\textit\def\PYG@tc##1{\textcolor[rgb]{0.24,0.48,0.48}{##1}}}
\@namedef{PYG@tok@cpf}{\let\PYG@it=\textit\def\PYG@tc##1{\textcolor[rgb]{0.24,0.48,0.48}{##1}}}
\@namedef{PYG@tok@c1}{\let\PYG@it=\textit\def\PYG@tc##1{\textcolor[rgb]{0.24,0.48,0.48}{##1}}}
\@namedef{PYG@tok@cs}{\let\PYG@it=\textit\def\PYG@tc##1{\textcolor[rgb]{0.24,0.48,0.48}{##1}}}

\def\PYGZus{\char`\_}
\def\PYGZob{\char`\{}
\def\PYGZcb{\char`\}}

\def\PYGZam{\char`\&}
\def\PYGZlt{\char`\<}
\def\PYGZgt{\char`\>}

\def\PYGZpc{\char`\%}

\def\PYGZhy{\char`\-}
\def\PYGZsq{\char`\'}
\def\PYGZdq{\char`\"}

\makeatother

\begin{document}

\title{Accelerating Accurate Assignment Authoring Using Solution-Generated Autograders}

\author{Geoffrey Challen}
\affiliation{%
  \institution{University of Illinois}
  \city{Urbana}
  \country{United States}
}
\email{challen@illinois.edu}

\author{Ben Nordick}
\affiliation{%
  \institution{Code Awakening LLC}
  \city{Champaign}
  \country{United States}
}
\email{ben@codeawakening.com}

\makeatletter
\gdef\@copyrightpermission{
  \begin{minipage}{\columnwidth}
  \vspace{3pt}
  \begin{minipage}{0.3\columnwidth}
  \href{https://creativecommons.org/licenses/by/4.0/}{\includegraphics[width=0.90\columnwidth]{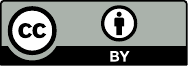}}
  \end{minipage}
  \hfill
  \begin{minipage}{0.7\columnwidth}
\href{https://creativecommons.org/licenses/by/4.0}{Work is licensed under a Creative Commons Attribution International 4.0 License.}
  \end{minipage}
  \vspace{3pt}
  \end{minipage}
}
\makeatother

\renewcommand{\shortauthors}{Geoffrey Challen and Ben Nordick}

\newcommand{\FIXME}[1]{\textbf{\textcolor{red}{#1}}}
\newcommand{\Questioner}{\textsc{Questioner}}
\newcommand{\SnapAct}{\textsc{Snapact\hyphenation{Snap-act}}}
\newcommand{\Jeed}{\textsc{Jeed}}
\newcommand{\learncsonline}{\texttt{learncs.online}}
\newcommand{\classidentifier}[1]{\textbf{\texttt{\textcolor{blue}{#1}}}}
\newcommand{\methodidentifier}[1]{\texttt{\textcolor{blue}{#1}}}

\begin{abstract}

Students learning to program benefit from access to large numbers of practice problems.
Autograders are commonly used to support programming questions by providing quick feedback on submissions.
But authoring accurate autograders remains challenging.
Autograders are frequently created by enumerating test cases---a tedious process that can produce inaccurate autograders that fail to correctly classify submissions.
When authoring accurate autograders is slow, it is difficult to create large banks of practice problems to support beginning programmers.

We present \textit{solution-generated autograding}: a faster, more accurate, and more enjoyable way to create autograders.
Our approach leverages a key difference between software testing and autograding: The question author can provide a solution.
By starting with a solution, we can eliminate the need to manually enumerate test cases, validate the autograder's accuracy, and evaluate other aspects of submission code quality beyond behavioral correctness.
We describe \Questioner{}, an implementation of solution-generated autograding for Java and Kotlin, and share experiences from four years using \Questioner{} to support a large CS1 course: authoring nearly 800 programming questions used by thousands of students to evaluate millions of submissions.

\end{abstract}





\hyphenation{auto-grading}
\hyphenation{auto-grader}

\maketitle

\section{Introduction and Motivation}

Students learning to program benefit from practice solving many different programming challenges.
This makes authoring and maintaining large banks of programming problems beneficial for introductory computing courses.
Large question banks are advantageous in many ways.
They enable frequent small assessment, allowing regular low-stakes evaluation of students' programming abilities.
With enough questions, students might be assigned a few to solve on their own each week for credit, and then asked to solve several more on a weekly computer-based quiz.
With enough questions, that quiz can be randomized to improve security.
And with enough questions, some can be provided as worked examples, while still others can be available for reinforcement and additional practice.

Autograding---the automated assessment of coding challenge submissions---is important to make large problem banks usable.
Prompt and accurate human grading of submissions to hundreds of problems is infeasible.
Immediate autograder feedback allows students to work through multiple problems, learning as they go, without repeating the same mistakes.
As a result, authoring and maintaining large question banks means authoring and maintaining large numbers of autograders: one per question.

Authoring an autograder is usually done by writing a test suite, using existing language-specific testing frameworks such as JUnit for Java or pytest for Python.
The autograder then comprises a series of test cases run on student submissions to evaluate correctness.
But this common approach has multiple weaknesses.
First, an accurate test suite can require many test cases, and enumerating them can be tedious---particularly when the problem handles many corner cases or when the inputs or outputs are difficult to notate.
Second, the accuracy of the autograder can be difficult to assess.
We define autograder accuracy as its ability to distinguish between correct and incorrect submissions.
A small number of test cases may achieve accuracy for some problems, while other problems many require many more.
Feedback on autograder accuracy is rarely available, making it unclear how many test cases are needed, or even if the current ones are consistent with the intended solution.
Without this feedback, achieving high accuracy requires the ability to anticipate common programming mistakes.
Experienced instructors may be able to, but less-experienced staff may not.

Finally, testing frameworks are not designed for autograding and are not always well-suited to the task.
For example, errors caused by naming mistakes will typically be located in the test suite, not the submission, causing beginners to think the problem is with the test suite and not their own code.
Test suites are also not designed to be used blind---without access to the tests themselves.
Understanding the series of calls that led to a failure can be essential in debugging.
So when the autograding platform conceals the tests being used, this can result in student confusion and frustration.
There are times when an instructor may want autograder output to be intentionally inscrutable---perhaps to encourage students to write their own tests in more advanced courses.
But this should be a pedagogical decision, not the result of an autograder limitation.

In addition to these limitations, implementing autograders by writing test suites fails to harness a key difference between software testing and autograding.
When testing software, the tests and implementation are mutually reinforcing: Either can be incorrect or incomplete, and neither represents a source of truth.
In contrast, the author of a programming problem can provide a solution, which we refer to as the \textit{reference solution}.
The reference solution represents a single source of truth: A submission is correct if it reproduces the reference solution behavior, and incorrect otherwise.
As a result, the reference solution can generate an autograder, an approach we call \textit{solution-generated autograding}.
Solution-generated autograding is faster and more fun than authoring test suites, and generates accurate autograders that provide rich feedback to students.

We present the design and implementation of \Questioner{}, solution-generated autograding for Java and Kotlin programming problems.
Starting with only a description and reference solution, \Questioner{} uses the reference solution first to generate a autograder, next to establish its accuracy, and finally to extract code quality metrics used to evaluate submissions (\S\ref{sec:description}).
We also share our experiences using \Questioner{} to support a large CS1 course (\S\ref{sec:experiences}).
Since 2020 we have authored almost 800 problems used throughout our CS1 course: as worked examples, on daily homework, on weekly quizzes, and for ungraded practice.
For students, \Questioner{} rapidly provides accurate feedback on correctness and code quality.
For instructors, \Questioner{} enables rapid authoring of programming problems, significantly reducing the time required to assemble a large question bank.
\section{Description}
\label{sec:description}

\begin{example}
\begin{Verbatim}[commandchars=\\\{\},fontsize=\scriptsize,numbers=left,xleftmargin=0.2in,numbersep=0.11in]
\PYG{c+cm}{/* Create a public class named Question with a single class method ... */}
\PYG{n+nd}{@Correct}\PYG{p}{(}\PYG{n}{name}\PYG{+w}{ }\PYG{o}{=}\PYG{+w}{ }\PYG{l+s}{\PYGZdq{}}\PYG{l+s}{Three Ints Increasing Strict}\PYG{l+s}{\PYGZdq{}}\PYG{p}{,}
\PYG{+w}{         }\PYG{n}{author}\PYG{+w}{ }\PYG{o}{=}\PYG{+w}{ }\PYG{l+s}{\PYGZdq{}}\PYG{l+s}{challen@illinois.edu}\PYG{l+s}{\PYGZdq{}}\PYG{p}{,}
\PYG{+w}{         }\PYG{n}{version}\PYG{+w}{ }\PYG{o}{=}\PYG{+w}{ }\PYG{l+s}{\PYGZdq{}}\PYG{l+s}{2024.7.0}\PYG{l+s}{\PYGZdq{}}\PYG{p}{)}
\PYG{k+kd}{public}\PYG{+w}{ }\PYG{k+kd}{class} \PYG{n+nc}{Question}\PYG{+w}{ }\PYG{p}{\PYGZob{}}
\PYG{+w}{  }\PYG{k+kd}{public}\PYG{+w}{ }\PYG{k+kd}{static}\PYG{+w}{ }\PYG{k+kt}{boolean}\PYG{+w}{ }\PYG{n+nf}{areIncreasingStrict}\PYG{p}{(}\PYG{k+kt}{int}\PYG{+w}{ }\PYG{n}{one}\PYG{p}{,}\PYG{+w}{ }\PYG{k+kt}{int}\PYG{+w}{ }\PYG{n}{two}\PYG{p}{,}\PYG{+w}{ }\PYG{k+kt}{int}\PYG{+w}{ }\PYG{n}{three}\PYG{p}{)}\PYG{+w}{ }\PYG{p}{\PYGZob{}}
\PYG{+w}{    }\PYG{k}{return}\PYG{+w}{ }\PYG{n}{two}\PYG{+w}{ }\PYG{o}{\PYGZgt{}}\PYG{+w}{ }\PYG{n}{one}\PYG{+w}{ }\PYG{o}{\PYGZam{}}\PYG{o}{\PYGZam{}}\PYG{+w}{ }\PYG{n}{three}\PYG{+w}{ }\PYG{o}{\PYGZgt{}}\PYG{+w}{ }\PYG{n}{two}\PYG{p}{;}
\PYG{+w}{  }\PYG{p}{\PYGZcb{}}
\PYG{p}{\PYGZcb{}}
\end{Verbatim}
\caption{\textbf{Example \Questioner{} Question.}}
\label{example:simple}
\end{example}

Example~\ref{example:simple} shows an example \Questioner{} question.
We begin by providing an description of the process of using only this information to generate an accurate autograder.
Next, we discuss situations where \Questioner{} requires additional information from the author.
Because \Questioner{} is a Java and Kotlin implementation of solution-generated autograding, we will present examples in Java and utilize JVM-specific terminology when appropriate.

We refer to the entire contents of Example~\ref{example:simple} as the \textit{control class}.
The control class always includes three key pieces of information: (1) the problem description, (2) configuration components used during autograder generation, and (3) the reference solution.
In Example~\ref{example:simple} above, the description is parsed from the comment on line 1.
The only required configuration is the \texttt{@Correct} annotation (lines 2--4), which defines the name, author, version triple that identifies the question.
%
% \Questioner{} uses this annotation to identify control files, and specifying question coordinates manually makes them stable across file renames.
%
The reference solution is extracted from the control class by removing the description and any configuration components, and in this case would comprise lines 5--9.

\subsection{Generating Inputs and Comparing Behavior}

Autograder generation begins by analyzing the reference solution to determine how to reveal its behavior.
In Example~\ref{example:simple}, \Questioner{} identifies that the reference solution class contains a single \texttt{static} method accepting three \texttt{int} parameters and returning a \texttt{boolean}.
To call this method, \Questioner{} does not need to create an instance of \texttt{Question} (the method is \texttt{static}), but does need to generate \texttt{int}s.
And to compare the behavior of the solution to a submission, \Questioner{} must be able to compare \texttt{boolean}s.

\Questioner{} includes built-in random generators for both single values and collections of all common Java types.
So it knows how to generate \texttt{int} values.
And it can compare any Java type for equality---manually for primitive types and via \texttt{.equals} for objects.
So it known how to compare \texttt{boolean} values.
We will discuss how and why input generation and output comparison can be customized, but this problem utilizes the default generators and comparators.

Note that \Questioner{} expects the submission to match the behavior of the reference solution in multiple ways.
Given the same history and inputs, a submission method should return a value equal to the same reference solution method when called with the same parameters.
But to be correct, a submission should also \texttt{throw} an exception of the same type when the reference solution throws, generate the same standard output or error, and manipulate state like the filesystem in the same way.
Overall, any publicly visible change make by the reference solution should be mirrored by the submission.
However, \Questioner{} places few requirements on \textit{how} the submission achieves the desired behavior, accepting any submission that mirrors the reference solution, and providing implementation flexibility to students.
We discuss some exceptions to this flexibility in Section~\ref{subsec:quality}.

\subsection{Establishing Accuracy}

At this point we have formulated an autograding strategy for Example~\ref{example:simple}.
\Questioner{} will generate triplets of random \texttt{int}s, use them as parameters to call the reference solution and submission \methodidentifier{areIncreasingStrict} method, and compare the results.

But how many times should \Questioner{} repeat this process?
Too few iterations may result in an inaccurate autograder.
Because we are comparing to the reference solution, submissions that match its behavior exactly will never be considered incorrect.
But there is still the possibility of erroneously classifying incorrect submissions as correct if they are not sufficiently examined.
On the other hand, performing too many iterations may cause autograding to become slow and computationally expensive.
Depending on the question, \Questioner{} can generate many potential inputs---$2^{96}$ in this case---making exhaustive exploration infeasible.

To generate an accurate autograder, \Questioner{} must use as many iterations as required to identify common mistakes students might make when attempting the problem.
If \Questioner{} had access to a collection of submissions known to be incorrect, it could use them to determine the number of iterations needed to achieve accuracy.
But \Questioner{} is generating the autograder used to classify submissions!
So assembling an incorrect corpus would require laborious human examination of student submissions before the autograder's accuracy could be established.

\begin{example}
\begin{Verbatim}[commandchars=\\\{\},fontsize=\scriptsize]
\PYG{k+kd}{public}\PYG{+w}{ }\PYG{k+kt}{boolean}\PYG{+w}{ }\PYG{n+nf}{areStrictlyIncreasing}\PYG{p}{(}\PYG{k+kt}{int}\PYG{+w}{ }\PYG{n}{one}\PYG{p}{,}\PYG{+w}{ }\PYG{k+kt}{int}\PYG{+w}{ }\PYG{n}{two}\PYG{p}{,}\PYG{+w}{ }\PYG{k+kt}{int}\PYG{+w}{ }\PYG{n}{three}\PYG{p}{)}\PYG{+w}{ }\PYG{p}{\PYGZob{}}
\PYG{+w}{  }\PYG{k}{return}\PYG{+w}{ }\PYG{n}{two}\PYG{+w}{ }\PYG{o}{\PYGZgt{}}\PYG{o}{=}\PYG{+w}{ }\PYG{n}{one}\PYG{+w}{ }\PYG{o}{\PYGZam{}}\PYG{o}{\PYGZam{}}\PYG{+w}{ }\PYG{n}{three}\PYG{+w}{ }\PYG{o}{\PYGZgt{}}\PYG{+w}{ }\PYG{n}{two}\PYG{p}{;}\PYG{+w}{ }\PYG{c+c1}{// conditional\PYGZhy{}boundary}
\PYG{p}{\PYGZcb{}}
\PYG{k+kd}{public}\PYG{+w}{ }\PYG{k+kt}{boolean}\PYG{+w}{ }\PYG{n+nf}{areStrictlyIncreasing}\PYG{p}{(}\PYG{k+kt}{int}\PYG{+w}{ }\PYG{n}{one}\PYG{p}{,}\PYG{+w}{ }\PYG{k+kt}{int}\PYG{+w}{ }\PYG{n}{two}\PYG{p}{,}\PYG{+w}{ }\PYG{k+kt}{int}\PYG{+w}{ }\PYG{n}{three}\PYG{p}{)}\PYG{+w}{ }\PYG{p}{\PYGZob{}}
\PYG{+w}{  }\PYG{k}{return}\PYG{+w}{ }\PYG{n}{two}\PYG{+w}{ }\PYG{o}{\PYGZlt{}}\PYG{o}{=}\PYG{+w}{ }\PYG{n}{one}\PYG{+w}{ }\PYG{o}{\PYGZam{}}\PYG{o}{\PYGZam{}}\PYG{+w}{ }\PYG{n}{three}\PYG{+w}{ }\PYG{o}{\PYGZgt{}}\PYG{+w}{ }\PYG{n}{two}\PYG{p}{;}\PYG{+w}{ }\PYG{c+c1}{// negate\PYGZhy{}conditional}
\PYG{p}{\PYGZcb{}}
\PYG{k+kd}{public}\PYG{+w}{ }\PYG{k+kt}{boolean}\PYG{+w}{ }\PYG{n+nf}{areStrictlyIncreasing}\PYG{p}{(}\PYG{k+kt}{int}\PYG{+w}{ }\PYG{n}{one}\PYG{p}{,}\PYG{+w}{ }\PYG{k+kt}{int}\PYG{+w}{ }\PYG{n}{two}\PYG{p}{,}\PYG{+w}{ }\PYG{k+kt}{int}\PYG{+w}{ }\PYG{n}{three}\PYG{p}{)}\PYG{+w}{ }\PYG{p}{\PYGZob{}}
\PYG{+w}{  }\PYG{k}{return}\PYG{+w}{ }\PYG{n}{three}\PYG{+w}{ }\PYG{o}{\PYGZgt{}}\PYG{+w}{ }\PYG{n}{two}\PYG{p}{;}\PYG{+w}{ }\PYG{c+c1}{// remove\PYGZhy{}and\PYGZhy{}or}
\PYG{p}{\PYGZcb{}}
\end{Verbatim}
\caption{\textbf{Example Mutants for Example~\ref{example:simple}.} \Questioner{} generates mutants from the reference solution to establish autograder accuracy. (The \texttt{Question} class wrapper has been omitted.)}
\label{example:mutants}
\end{example}

\Questioner{} addresses this challenge by using the reference solution itself to generate incorrect submissions via \textit{mutation}.
Starting with the reference solution, \Questioner{} uses mutation to introduce small changes that affect program behavior---such as replacing \texttt{<} with \texttt{<=} or swapping \texttt{\&\&} and \texttt{||}.
\Questioner{} utilizes a source-level mutation library containing 37 Java and Kotlin mutators included as part of the \Jeed{} pedagogical source code execution and analysis toolkit.
Source-level mutation is used so that mutated examples can be presented to question authors when autograder generation fails.
Example~\ref{example:simple} generates nine mutated incorrect submissions, several of which are shown in Example~\ref{example:mutants}.
Question authors may still manually augment the mutation-generated incorrect corpus, and may occasionally need to disable certain mutants for a question.
But our example uses only the mutants to establish accuracy.

Using the corpus of incorrect examples generated from the reference solution, \Questioner{} can now determine how many iterations are required to produce an accurate autograder.
\Questioner{} tests each incorrect example by comparing its behavior to the reference solution.
Mutants fail to match the reference solution when their behavior diverges or when testing times out---since mutation may introduce infinite loops.
To ensure reproducibility, the same sequence of random inputs is used each time.
If all incorrect examples are identified, the required number of iterations is determined, the accuracy of the autograder is established, and generation is complete.
For the question in Example~\ref{example:simple}, 21 iterations were required to identify the nine incorrect examples produced by mutation.

However, it is also possible that all incorrect examples are \textit{not} identified, at least not before \Questioner{} reaches a configurable limit on the number of iterations.
When autograder generation fails, \Questioner{} presents the misidentified example to the author who determines how to proceed.
This can happen for two reasons.
It is possible that code that was expected to be incorrect is actually correct.
For some problems, certain mutations produce identical behavior---for example, swapping \texttt{>} and \texttt{>=} when determining the maximum of an integer array.
To handle this case, \Questioner{} allows authors to suppress mutations through comments.
A single suppression is usually sufficient to allow autograder generation to succeed.
But it is also possible that the misidentified example is actually incorrect but \Questioner{} did not find an input that revealed the mistake.
We return to this scenario in Section~\ref{subsec:inputgeneration}.

\subsection{Constraints and Quality}
\label{subsec:quality}

At this point we have generated and validated an autograder.
However, there is still additional information that \Questioner{} obtains from the reference solution---both to constrain the autograding environment and to provide code quality feedback.
As an example constraint, \Questioner{} records the classes used by the reference solution during execution.
These are used to limit the classes available to submissions, an important capability to ensure that students meet pedagogical goals.
For example, if the challenge is to implement a sorting algorithm, \Questioner{} should not allow submissions to utilize Java's built-in sorting methods.
The control class may whitelist libraries not used by the reference solution to provide additional implementation flexibility as appropriate.

\begin{comment}
    
\begin{example}
\begin{Verbatim}[commandchars=\\\{\},fontsize=\scriptsize]
\PYG{k+kd}{public}\PYG{+w}{ }\PYG{k+kt}{boolean}\PYG{+w}{ }\PYG{n+nf}{areStrictlyIncreasing}\PYG{p}{(}\PYG{k+kt}{int}\PYG{+w}{ }\PYG{n}{one}\PYG{p}{,}\PYG{+w}{ }\PYG{k+kt}{int}\PYG{+w}{ }\PYG{n}{two}\PYG{p}{,}\PYG{+w}{ }\PYG{k+kt}{int}\PYG{+w}{ }\PYG{n}{three}\PYG{p}{)}\PYG{+w}{ }\PYG{p}{\PYGZob{}}
\PYG{+w}{  }\PYG{k}{if}\PYG{+w}{ }\PYG{p}{(}\PYG{n}{one}\PYG{+w}{ }\PYG{o}{=}\PYG{o}{=}\PYG{+w}{ }\PYG{n}{two}\PYG{+w}{ }\PYG{o}{|}\PYG{o}{|}\PYG{+w}{ }\PYG{n}{two}\PYG{+w}{ }\PYG{o}{=}\PYG{o}{=}\PYG{+w}{ }\PYG{n}{three}\PYG{p}{)}\PYG{+w}{ }\PYG{p}{\PYGZob{}}
\PYG{+w}{    }\PYG{k}{return}\PYG{+w}{ }\PYG{k+kc}{false}\PYG{p}{;}
\PYG{+w}{  }\PYG{p}{\PYGZcb{}}
\PYG{+w}{  }\PYG{k}{return}\PYG{+w}{ }\PYG{n}{two}\PYG{+w}{ }\PYG{o}{\PYGZgt{}}\PYG{+w}{ }\PYG{n}{one}\PYG{+w}{ }\PYG{o}{\PYGZam{}}\PYG{o}{\PYGZam{}}\PYG{+w}{ }\PYG{n}{three}\PYG{+w}{ }\PYG{o}{\PYGZgt{}}\PYG{+w}{ }\PYG{n}{two}\PYG{p}{;}
\PYG{p}{\PYGZcb{}}
\end{Verbatim}
\caption{\textbf{Unnecessarily Complex Solution to Example~\ref{example:simple}.} The example shown is correct, but has four code paths where only two are needed. (The \texttt{Question} class wrapper has been omitted.)}
\label{example:complexity}
\end{example}

\end{comment}

\Questioner{} also utilizes static analysis and runtime instrumentation to collect other information about the reference solution used to evaluate submission \textit{code quality}.
We define code quality as other aspects of a submission beyond correctness that indicate a good approach.
As one example, \Questioner{} records the \textit{cyclomatic complexity}~\cite{cyclomaticcomplexitywikipedia} of the reference solution---the number of code paths used to complete the problem.
Student submissions that use many more code paths can be simplified, and may indicate a conceptual misunderstanding.
After acknowledging correctness, autograder feedback can inform the student that their code can be simplified.

In addition to complexity, \Questioner{} currently measures the following aspects of the reference solution to evaluate code quality:

\begin{enumerate}
    \item \textbf{Execution time:} as the number of submission lines executed.
    While this does not directly translate to work done by the program, it is stable across machines and testing runs, unlike wall clock or processor execution time.
    \item \textbf{Memory usage:} as the total amount of memory allocated.
    \item \textbf{Submission length:} as the count of non-commenting lines.
    \item \textbf{Dead code:} any lines not fully executed during testing.
    \item \textbf{Recursive method implementations:} identified using bytecode instrumentation.
\end{enumerate}

All the code quality aspects evaluated by \Questioner{} are motivated by our experiences teaching beginning programmers to write better code.
By default, student submissions are not expected to exactly match the code quality metrics measured from the reference solution, but large deviations may indicate an opportunity to provide helpful feedback.
For example, a submission that executes 50\% more lines than the reference solution may be fine---but one that executes 4 times as many may be worth fixing.
Instructors may always configure how to utilize code quality information during autograding---both whether code quality should be graded, and whether code quality hints should be shown to students.

How code quality metrics collected are used varies.
For the efficiency metrics (execution time, memory usage, submission length), the control class can configure how tightly submission must match the reference solution, with tight bounds supporting programming challenges focused on performance.
For recursive methods, if \Questioner{} identifies that a reference solution method was implemented recursively, submissions must also implement that method recursively, ensuring that students don't submit iterative solutions when a recursive approach was requested.
For dead code, any dead code present in the reference solution will cause the problem to fail autograder generation.
If needed, problem authors can increase the question dead code limit either through control class configuration or via end-of-line comments.
This allows handling cases such as when a \texttt{return} inside a loop will always be reached, and as a result the function-ending \texttt{return} will be marked as dead.

Finally, \Questioner{} also uses complexity to defend the autograder.
One concern with autograders is that frustrated students may submit brute-force implementations that enumerate all test cases inside an \texttt{if-else} statement rather than implement a solution for all inputs.
%
% If the number of test cases is small and the output provided descriptive this may be a viable approach, and we have heard students confirm that it works for certain courses, even during high-stakes computerized assessments.
%
\Questioner{} autograders typically utilize many more inputs than an instructor would by hand, but we have still seen students attempt this strategy.
To ensure it does not succeed, \Questioner{} sets a hard limit on submission complexity that is both much higher than the reference solution and much lower than the number of iterations used by the autograder.
Submissions exceeding this threshold are immediately rejected.

\begin{example}
\begin{Verbatim}[commandchars=\\\{\},fontsize=\scriptsize]
\PYG{k+kd}{public}\PYG{+w}{ }\PYG{k+kd}{class} \PYG{n+nc}{Question}\PYG{+w}{ }\PYG{p}{\PYGZob{}}
\PYG{+w}{  }\PYG{n+nd}{@FixedParameters}
\PYG{+w}{  }\PYG{k+kd}{private}\PYG{+w}{ }\PYG{k+kd}{static}\PYG{+w}{ }\PYG{k+kd}{final}\PYG{+w}{ }\PYG{n}{List}\PYG{o}{\PYGZlt{}}\PYG{n}{Integer}\PYG{o}{\PYGZgt{}}\PYG{+w}{ }\PYG{n}{FIXED}\PYG{+w}{ }\PYG{o}{=}\PYG{+w}{ }\PYG{n}{Arrays}\PYG{p}{.}\PYG{n+na}{asList}\PYG{p}{(}\PYG{l+m+mi}{88}\PYG{p}{,}\PYG{+w}{ }\PYG{l+m+mi}{888}\PYG{p}{)}\PYG{p}{;}
\PYG{+w}{  }\PYG{k+kd}{public}\PYG{+w}{ }\PYG{k+kd}{static}\PYG{+w}{ }\PYG{k+kt}{boolean}\PYG{+w}{ }\PYG{n+nf}{isSecretNumber}\PYG{p}{(}\PYG{k+kt}{int}\PYG{+w}{ }\PYG{n}{value}\PYG{p}{)}\PYG{+w}{ }\PYG{p}{\PYGZob{}}
\PYG{+w}{    }\PYG{k}{return}\PYG{+w}{ }\PYG{n}{value}\PYG{+w}{ }\PYG{o}{=}\PYG{o}{=}\PYG{+w}{ }\PYG{l+m+mi}{88}\PYG{+w}{ }\PYG{o}{|}\PYG{o}{|}\PYG{+w}{ }\PYG{n}{value}\PYG{+w}{ }\PYG{o}{=}\PYG{o}{=}\PYG{+w}{ }\PYG{l+m+mi}{888}\PYG{p}{;}
\PYG{+w}{  }\PYG{p}{\PYGZcb{}}
\PYG{p}{\PYGZcb{}}
\end{Verbatim}
\caption{\textbf{Question Requiring Special Inputs.} Two special inputs are provided using the \texttt{@FixedParameters} annotation.}
\label{example:special}
\end{example}

\subsection{Custom Input Generators}
\label{subsec:inputgeneration}

For Example~\ref{example:simple} \Questioner{}'s default input generators successfully generate an accurate autograder.
But this is not always the case.
Input generation can fail for two reasons: either because \Questioner{} does not know how to generate inputs of a particular type, or because the default inputs of that type do not achieve accuracy.

As an example of the first case, if the reference solution method expects an instance of \texttt{class Foo}, \classidentifier{Foo} being a custom class utilized by this problem, \Questioner{} will not know how to generate inputs of type \classidentifier{Foo}.
Example~\ref{example:special} shows an example of the second case.
\Questioner{} knows how to generate random \texttt{int} values, but is unlikely to uncover the special values needed by this question.
Autograder generation for this problem will fail, because several mutants will not be correctly identified.

\Questioner{} handles both of these cases by allowing authors to configure custom input generation on the control class: either by providing a list of inputs, a method to generate inputs, or both.
A list works well when the default generator is nearly correct but a small number of special values must be included, as in the problem shown in Example~\ref{example:special}.
A generator method handles cases where inputs must have some special property that the default random generator does not produce---for example, if the challenge is to identify palindromic \texttt{String}s, and a high percentage of inputs should be palindromes.
The same mechanism also allows authors to generate input types that \Questioner{} is unfamiliar with.
%
% Both custom input lists and generator methods are marked with \Questioner{} annotations, allowing them to be identified during autograder generation and removed when extracting the reference solution.

It may appear that we have resorted to enumerating test cases---the tedious process that solution-generated autograding was supposed to avoid.
However, \Questioner{}'s input customization mechanisms are both more compact and more powerful than authoring test suites.
It is almost always easier to generate special inputs programmatically rather than notating them individually.
For example, an infinite number of random palindromes can be generated by creating a random \texttt{String} and appending its reverse.
And \Questioner{} still utilizes the reference solution to determine correct behavior, avoiding the mistakes possible when providing input-output pairs.
Finally, custom input generators still harness all of the autograder generation capabilities previously described: \Questioner{} still uses incorrect examples to determine how many iterations are required to produce an accurate autograder.
Note that \Questioner{} can choose an iteration count precisely because an input generation method produces a potentially-infinite series of inputs, compared with the fixed number represented by a set of test cases.

\begin{example}
\begin{Verbatim}[commandchars=\\\{\},fontsize=\scriptsize]
\PYG{k+kd}{public}\PYG{+w}{ }\PYG{k+kd}{class} \PYG{n+nc}{EvenOddFlipFlop}\PYG{+w}{ }\PYG{p}{\PYGZob{}}
\PYG{+w}{  }\PYG{k+kd}{private}\PYG{+w}{ }\PYG{k+kt}{boolean}\PYG{+w}{ }\PYG{n}{isEven}\PYG{p}{;}
\PYG{+w}{  }\PYG{k+kd}{public}\PYG{+w}{ }\PYG{n+nf}{EvenOddFlipFlop}\PYG{p}{(}\PYG{k+kt}{int}\PYG{+w}{ }\PYG{n}{setValue}\PYG{p}{)}\PYG{+w}{ }\PYG{p}{\PYGZob{}}
\PYG{+w}{    }\PYG{n}{isEven}\PYG{+w}{ }\PYG{o}{=}\PYG{+w}{ }\PYG{n}{setValue}\PYG{+w}{ }\PYG{o}{\PYGZpc{}}\PYG{+w}{ }\PYG{l+m+mi}{2}\PYG{+w}{ }\PYG{o}{=}\PYG{o}{=}\PYG{+w}{ }\PYG{l+m+mi}{0}\PYG{p}{;}
\PYG{+w}{  }\PYG{p}{\PYGZcb{}}
\PYG{+w}{  }\PYG{k+kd}{public}\PYG{+w}{ }\PYG{k+kt}{void}\PYG{+w}{ }\PYG{n+nf}{setValue}\PYG{p}{(}\PYG{k+kt}{int}\PYG{+w}{ }\PYG{n}{setValue}\PYG{p}{)}\PYG{+w}{ }\PYG{p}{\PYGZob{}}
\PYG{+w}{    }\PYG{k}{assert}\PYG{p}{(}\PYG{o}{!}\PYG{p}{(}\PYG{n}{isEven}\PYG{+w}{ }\PYG{o}{\PYGZam{}}\PYG{o}{\PYGZam{}}\PYG{+w}{ }\PYG{n}{setValue}\PYG{+w}{ }\PYG{o}{\PYGZpc{}}\PYG{+w}{ }\PYG{l+m+mi}{2}\PYG{+w}{ }\PYG{o}{=}\PYG{o}{=}\PYG{+w}{ }\PYG{l+m+mi}{0}\PYG{p}{)}\PYG{p}{)}\PYG{p}{;}
\PYG{+w}{    }\PYG{n}{isEven}\PYG{+w}{ }\PYG{o}{=}\PYG{+w}{ }\PYG{n}{setValue}\PYG{+w}{ }\PYG{o}{\PYGZpc{}}\PYG{+w}{ }\PYG{l+m+mi}{2}\PYG{+w}{ }\PYG{o}{=}\PYG{o}{=}\PYG{+w}{ }\PYG{l+m+mi}{0}\PYG{p}{;}
\PYG{+w}{  }\PYG{p}{\PYGZcb{}}
\PYG{+w}{  }\PYG{k+kd}{public}\PYG{+w}{ }\PYG{k+kt}{boolean}\PYG{+w}{ }\PYG{n+nf}{isEven}\PYG{p}{(}\PYG{p}{)}\PYG{+w}{ }\PYG{p}{\PYGZob{}}
\PYG{+w}{    }\PYG{k}{return}\PYG{+w}{ }\PYG{n}{isEven}\PYG{p}{;}
\PYG{+w}{  }\PYG{p}{\PYGZcb{}}
\PYG{p}{\PYGZcb{}}
\end{Verbatim}
\caption{\textbf{Example Class Design Question.} Solution-generated autograding works equally well for class design problems.}
\label{example:class}
\end{example}

\subsection{Class Design Questions}

So far we have presented examples of method-design problems---even if Java requires a \texttt{class} wrapper.
But solution-generated autograding works equally-well for class design problems.
Example~\ref{example:class} shows the control class for a class design question where updates to a stored \texttt{int} value are expected to alternate between even and odd.
Again, this is all an author needs to provide for \Questioner{} to generate an accurate autograder.

Generating autograders for class design questions proceeds similarly to what has been previously described.
Example~\ref{example:class} has a constructor and method that need \texttt{int} inputs, but for both the default random input generation is sufficient.
Unlike when testing classes that only contain \texttt{static} methods, \Questioner{} will create multiple instances of the \classidentifier{EvenOddFlipFlop} class, each initialized with different values passed to the constructor.
Java constructors do not return and are expected to \texttt{throw} on failure.
If both the reference solution and submission constructors \texttt{throw} the same kind of exception on the same inputs, \Questioner{} will continue by attempting to generate new instances using new constructor parameters.
Once a constructor call succeeds, \Questioner{} calls \classidentifier{EvenOddFlipFlop} methods in a random sequence passing them random inputs.

When classes maintain state, calling an incorrect method may not immediately reveal failure---and some instance methods may not even return a value.
For example, a faulty implementation of \methodidentifier{setValue} for Example~\ref{example:class} may \texttt{assert} properly once but not update internal state correctly.
In this case, the failure will not be visible until the subsequent call to either \methodidentifier{setValue} or \methodidentifier{isEven}.
To facilitate debugging, when testing classes \Questioner{} feedback includes the entire failing sequence, not just the final call.

\begin{comment}

To generate an accurate autograder, the question class must provide visibility into private state manipulated by public methods.
%
% How this is done varies from question to question---in the example above, because only the evenness of \methodidentifier{value} is publicly exposed via \methodidentifier{isEven}, there is no need to save its exact value, a source of implementation flexibility for students.
%
For example, if a class provides a setter but no getter, \Questioner{} has no way to determine whether the setter is correct.
%
Questions with methods that include unobservable effects will typically fail autograder generation, when mistakes introduced by mutation affecting unobservable effects are not identified.

\end{comment}

\subsection{Additional Capabilities}
\label{subsec:capabilities}

\Questioner{} has many additional capabilities supporting accurate question authoring.
We briefly discuss several below.

\textbf{Output Comparison.}
By default method return values are compared for equality, but the control class can customize this comparison if needed.
For example, Java's \methodidentifier{Comparable<T>} interface specifies only the sign of the \methodidentifier{compareTo} return, not its exact value.

\textbf{Method and Snippet Variants.}
Depending on the pedagogical approach used by the course, students may not be immediately expected to author complete classes.
\Questioner{} allows questions that accept either only a method or a snippet of code that would normally appear inside a method.
Submissions to these questions are inserted into a template extracted from the control class, at which point autograding continues as usual.

\textbf{Starter Code.}
\Questioner{} problems can provide students with starter code when appropriate.
Starter code is included with other incorrect examples used during autograder generation to ensure that it does not accidentally solve the problem.

\textbf{Kotlin Support.}
\Questioner{} can autograde Kotlin submissions to questions configured using Java control classes.
To add Kotlin support, the author provides a Kotlin solution and description.
A separate description is required because Kotlin has different syntax and idioms.
During autograder generation, \Questioner{} ensures that the Kotlin solution is correct and extracts separate code quality metrics to evaluate Kotlin submissions.

\textbf{Feature Analysis.}
\Questioner{} uses static analysis to identify the language syntax features used by the reference solution.
This information allows a large problem bank to be cataloged (see Table~\ref{table:features}) and students to identify problems that provide practice with a given concept.
The control class can also examine submission features during grading, allowing it to enforce requirements such as not to use looping constructs.

\subsection{Tooling}
\label{subsec:tooling}

\Questioner{} provides a Gradle plugin to support problem authoring, allowing authors to work in an IDE.
The plugin orchestrates the process of autograder generation, providing feedback to authors both through error messages and through HTML reports, and allows authors to publish questions to a \Questioner{} backend.

The \Questioner{} backend is a Dockerized JVM server application providing a JSON web API.
Autograders for available questions are loaded from a MongoDB database.
The \Questioner{} backend API is unsecured.
In production, a separate public-facing proxy is used to authenticate requests, save submissions and autograding results, and update student scores when appropriate.

\begin{table*}[t]
{\fontsize{7.6pt}{8.6pt}\selectfont
\begin{tabularx}{0.20\textwidth}{Xr}
Declaring methods & 653 \\
Comparisons & 618 \\
\texttt{if-else} statements & 515 \\
Variable operations & 470 \\
Reference equality & 434 \\
Modifying variables & 431 \\
Declaring variables & 398 \\
Dotted notation & 393 \\
Initializing variables & 390 \\
\texttt{public} and \texttt{private} & 349 \\
\end{tabularx}%
\begin{tabularx}{0.20\textwidth}{Xr}
Declaring classes & 347 \\
Logical operators & 309 \\
Working with \texttt{null} & 296 \\
\texttt{for} and \texttt{while} loops & 268 \\
Strings & 254 \\
Using arrays & 181 \\
\texttt{static} methods & 175 \\
Class field & 161 \\
Creating objects & 152 \\
\texttt{assert} statements & 138 \\
\end{tabularx}%
\begin{tabularx}{0.20\textwidth}{Xr}
Constructors & 134 \\
Throwing exceptions & 124 \\
Type parameters & 117 \\
\texttt{final} & 105 \\
\texttt{import} statements & 96 \\
Printing & 91 \\
Variable assignment & 91 \\
Enhanced \texttt{for} loop & 78 \\
Boxing classes & 76 \\
Equality & 76 \\
\end{tabularx}%
\begin{tabularx}{0.20\textwidth}{Xr}
Getters and setters & 71 \\
Recursion & 57 \\
\texttt{instanceof} checking & 57 \\
Binary trees & 47 \\
Lists & 38 \\
Extending classes & 36 \\
Maps & 34 \\
Implementing interfaces & 29 \\
Nested conditionals & 28 \\
Nested loops & 27 \\
\end{tabularx}%
\begin{tabularx}{0.20\textwidth}{Xr}
Sets & 24 \\
Calling \texttt{super} & 18 \\
\texttt{break} or \texttt{continue} & 14 \\
Object type casting & 9 \\
Graphs & 7 \\
Lambda expressions & 7 \\
\texttt{try-catch} & 6 \\
Inner classes & 4 \\
Primitive type casting & 3 \\
Anonymous classes & 3 \\
\end{tabularx}%
}

\vspace{2pt}
\caption{\textbf{Java Features Used by Problems in Our 771~Question CS1 \Questioner{} Problem Bank.}
We have successfully used \Questioner{} to author problems testing all core CS1 programming concepts.
Note that problems almost always use more than one feature.
}
\label{table:features}
\end{table*}

\subsection{Limitations}
\label{subsec:limitations}

Currently \Questioner{} only compares the behavior of one JVM class per problem, meaning that all tested methods must be provided by that class.
\Questioner{} questions can still use multiple classes, but any additional classes are shared between the reference solution and submission and not authored by the student.
At times this requires some reorganization of existing problems when adapting them for \Questioner{} support.
Note that this is a limitation of our implementation, and not of solution-generated autograding.
\SnapAct{}, our newer implementation of solution-generated autograding for Python problems, removes this limitation.

\Questioner{} does not attempt to generate minimal autograders, meaning that it may use inputs that do not identify mistakes and more test cases than a skilled human developer.
While pruning unneeded iterations is technically feasible and would improve performance, \Questioner{} autograding already completes within 100s of milliseconds, reducing the need for further performance tuning.

\Questioner{} may incorrectly evaluate adversarial submissions---for example, an \texttt{int addOne(int)} method that correctly adds one to its argument, except when the argument is \methodidentifier{960950} or some other random \texttt{int}.
Autograders implemented using test suites also share this limitation.
\Questioner{} can identify the extra code paths used to achieve this behavior, and could potentially reject adversarial submissions.
However, this kind of submission typically indicates a student who has met the pedagogical goals of the exercise.

% While \Questioner{} supports both Java and Kotlin, question control classes must be authored in Java due to the parsing needed to extract the reference solution.
\section{Experiences}
\label{sec:experiences}

\begin{figure}[t]
\includegraphics[width=\columnwidth]{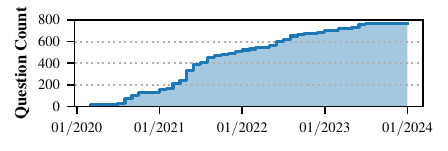}
\caption{\textbf{Growth of Our CS1 \Questioner{} Problem Bank.}}
\label{figure:growth}
\Description[Growth of our CS1 \Questioner{} problem bank.]{Our CS1 \Questioner{} problem bank has grown to 771 questions in roughly three years.}
\end{figure}

Since Fall 2020 we have used \Questioner{} to generate autograders for programming questions used by \href{https://cs124.org}{CS 124}, a large CS1 course at the University of Illinois.
Our experience demonstrates how solution-generated autograding accelerates question authoring.

Figure~\ref{figure:growth} plots the growth of our problem bank, which currently contains 771~questions.
We have been able to use \Questioner{} to support problems assessing all core CS1 language features, as shown in Table~\ref{table:features}.
Students may choose to take the course in either Java or Kotlin, and 742~questions (96\%) support both languages.

\Questioner{} problems are integrated into the course in multiple ways.
Students complete 51~questions as graded homework, assigned daily during most weeks.
103~questions appear on daily lessons as practice problems with worked solutions.
Students take 15~weekly quizzes, which together include 39~question slots, each of which randomly selects one from several similar problems.
176~questions are used on graded quizzes, an average of 4.5~per question.
An additional 110 appear on practice quizzes.
The remaining 331~questions (43\%) are made available to students for additional practice.
Students submit to \Questioner{} and receive feedback through a online interface for homework, practice, and quiz questions.

While course staff have recently begun to author questions, the 771~question bank described here was authored by a single instructor, who was the sole instructor responsible for the course during all but one term.
Given that most questions were written during the three years between 8/2020 and 8/2023, a single instructor was able to author approximately one question per working day over that time period.
Each fall the goal was to write a new set of problems for most of the daily lessons, either additional practice or graded homework exercises.
And every semester the goal was to write a new set of questions for each weekly quiz.
\Questioner{}'s implementation of solution-generated autograding allowed these goals to be met in a few hours per week, leaving ample time for other course management and delivery tasks.
During the 2023--2024 academic year, with a sizable problem bank in place, new question development slowed.
Going forward, the focus will be on training course staff to write questions using \Questioner{}.

In Spring~2024 \Questioner{} received \num{851192} homework and practice problem submissions from \num{1054} users.
\num{815} students enrolled in the course that term, but several hundred course staff also generate submissions.
\num{850881} submissions were successfully graded (99.96\%), with a small number of failures caused by backend restarts.
Four \Questioner{} backend servers were utilized, with one pair handling quiz submissions and the second pair homework and practice submissions.
The median time to evaluate submissions across all questions was 43~ms and the 99th~percentile was 806~ms, enabling interactive use by students.
Details of how this performance is achieved are beyond the scope of this paper, but \Questioner{} leverages high-performance compilation and sandboxed execution capabilities provided by the \Jeed{} library.
Students also submitted \num{748937} answers to quiz programming questions.

Autograder accuracy is difficult to establish based on student reports, given that they will complain vociferously when code they consider correct is rejected but tend to be quieter when incorrect code is accepted.
\Questioner{}'s use of the solution to establish correct behavior does help ensure that correct submissions are accepted, important when utilizing autograders in higher-stakes environments like quizzes.
For a typical quiz, new questions are authored several days beforehand and reviewed by a handful of senior course staff to ensure the description matches the solution.
Despite authoring hundreds of new quiz questions since 2020, we have never had to drop a quiz question due to autograder inaccuracy.

Starting in 2023, two external instructors have utilized \Questioner{} to author 59~questions and publish them to \learncsonline{}, a public website hosting introductory computer science materials.
Guided by a tutorial, they have been able to quickly use \Questioner{} to generate accurate autograders, including successfully adjusting problem configurations as needed when generation fails.
\section{Related Work}
\label{sec:related}

The benefits of providing students with immediate feedback has led to widespread use of autograding and the development of many autograding platforms~\cite{lin2020berkeley}.
We use the term autograding platform to distinguish the autograder itself---typically provided by the question author---from the software that receives submissions, runs the autograder, updates the gradebook, and displays results to students.

\begin{comment}
The autograding platforms we have examined usually support one of two approaches.
%
Container-based autograding platforms execute an instructor-provided container, which is provided with the submission and possibly other question-specific information, and is expected to return structured grading and feedback data to the platform before exiting.
%
PrairieLearn's external grader~\cite{west2015prairielearn,prairielearnurl} and Gradescope's autograder~\cite{singh2017gradescope,gradescopeurl} are two examples of this approach.
%
Testing-based autograding platforms expect instructors to implement autograders by enumerating test cases---either in code using an appropriate testing library (e.g., WebCat~\cite{shah2003web,webcaturl}), provided through a web interface (e.g., CodeGrade~\cite{codegradeurl}), or in a bespoke format (e.g., CodingBat~\cite{kiesler2023investigating,codingbaturl}).
%
Neither type of autograding platform assists with autograder authoring, and testing-based platforms reflect the assumption that enumerating test cases is necessary.
\end{comment}
The autograding platforms we have examined require the instructor to provide artifacts beyond or in place of a solution.
Many require enumerating test cases---either in code using an appropriate testing library (e.g., WebCat~\cite{shah2003web,webcaturl}), provided through a web interface (e.g., CodeGrade~\cite{codegradeurl}), or in a bespoke format (e.g., CodingBat~\cite{kiesler2023investigating,codingbaturl}).
Other platforms leave the process of invoking the student code to the instructor, requiring an instructor-provided container to output structured grading and feedback data given a submission.
PrairieLearn's external grader~\cite{west2015prairielearn,prairielearnurl} and Gradescope's autograder~\cite{singh2017gradescope,gradescopeurl} are two examples of this approach.
Neither type of autograding platform assists with autograder authoring, and testing-based platforms reflect the assumption that enumerating test cases is necessary.
We are not aware of another implementation of solution-generated autograding, but it is hard to gather information about the many bespoke autograders in use throughout computer science programs.

Solution-generated autograding shares some similarities with property-based testing~\cite{fink1997property}, an approach that verifies output properties rather than output equality.
But property-based testing is not appropriate for all situations, and still requires test authors to specify properties---whereas solution-generated autograding automatically extracts required properties from the reference solution.
\section{Future Work}
\label{sec:futurework}

\begin{example}
\begin{Verbatim}[commandchars=\\\{\},fontsize=\scriptsize]
\PYG{n}{QUESTION}\PYG{p}{(}\PYG{n}{name}\PYG{o}{=}\PYG{l+s+s1}{\PYGZsq{}}\PYG{l+s+s1}{Add One}\PYG{l+s+s1}{\PYGZsq{}}\PYG{p}{,} \PYG{n}{version}\PYG{o}{=}\PYG{l+s+s1}{\PYGZsq{}}\PYG{l+s+s1}{1.0}\PYG{l+s+s1}{\PYGZsq{}}\PYG{p}{,} \PYG{n}{author}\PYG{o}{=}\PYG{l+s+s1}{\PYGZsq{}}\PYG{l+s+s1}{ben@codeawakening.com}\PYG{l+s+s1}{\PYGZsq{}}\PYG{p}{)}
\PYG{l+s+sd}{\PYGZsq{}\PYGZsq{}\PYGZsq{} Write a function called `add\PYGZus{}one`... \PYGZsq{}\PYGZsq{}\PYGZsq{}}
\PYG{k}{def}\PYG{+w}{ }\PYG{n+nf}{add\PYGZus{}one}\PYG{p}{(}\PYG{n}{i}\PYG{p}{:} \PYG{n+nb}{int}\PYG{p}{)} \PYG{o}{\PYGZhy{}}\PYG{o}{\PYGZgt{}} \PYG{n+nb}{int}\PYG{p}{:}
    \PYG{k}{return} \PYG{n}{i} \PYG{o}{+} \PYG{l+m+mi}{1}
\end{Verbatim}
\caption{\textbf{Example \SnapAct{} Question.}}
\label{example:snapact}
\end{example}

Solution-generated autograding can support any programming language, although support for static and dynamic analysis simplifies implementations.
We are currently completing work on \SnapAct{}, an implementation of solution-generated autograding for Python.
\SnapAct{} follows \Questioner{}'s approach while responding to unique challenges posed by Python.
For example, to determine how to generate inputs for Python methods, authors are required to utilize Python's support for type hints.
Enforcement of type annotations and consistency in submissions is supported but optional.
Example~\ref{example:snapact} shows an example \SnapAct{} question.

Solution-generated autograding can not only evaluate solutions, but also evaluate test suites.
To be correct, a student-authored test suite must accurately distinguish the reference solution from the incorrect examples created by mutation.
Note that test suite evaluation utilizes the same mutants used during autograder validation, requiring no additional work by the question author.
We began piloting these test-evaluation questions in CS~124 in Fall 2024.

We are also exploring how generative AI can combine with solution-generated autograding to further accelerate problem development.
After confirming the consistency of an LLM-generated description and reference solution, a problem author can use solution-generated autograding to quickly produce an accurate autograder.

\clearpage
\bibliographystyle{ACM-Reference-Format}
\bibliography{main/ref}
\end{document}